\documentstyle[amsfonts,twocolumn,aps,graphicx]{revtex}

\draft

\begin{document}
\title{Decoupling the $SU(N)_{2}$-homogeneous Sine-Gordon model}
\author{O.A.~Castro-Alvaredo$^\sharp$ and A.~Fring$\,^\star$}
\address{$^{\sharp }$Departamento de F\'{\i }sica de Part\'{\i }culas, Universidad
de Santiago de Compostela, E-15706 Santiago de Compostela, Spain\\
$^{\star }$Institut f\"{u}r Theoretische Physik, Freie Universit\"{a}t
Berlin, Arnimallee 14, D-14195 Berlin, Germany }
\date{\today}
\maketitle

\begin{abstract}
We provide a systematic construction for all $n$-particle form factors of
the $SU(N)_2/U(1)^{N-1}$-homogeneous Sine-Gordon model in terms of general
determinant formulae for a huge class of local operators. The ultraviolet
limit is carried out and the corresponding Virasoro central charge together
with the conformal dimensions of various operators are identified. The
renormalization group flow is studied and we find a precise rule, depending
on the relative order of magnitude of the resonance parameters, according to
which the theory decouples into new cosets along the flow.
\end{abstract}

\pacs{PACS numbers: 11.10Hi, 11.10Kk, 11.30Er, 05.70.Jk}


\section{Introduction}

For most integrable quantum field theories in 1+1 space-time dimensions it
remains an open challenge to complete the entire bootstrap program, i.e. to
compute the exact on-shell S-matrix, closed formulae for the $n$-particle
form factors, identify the entire local operator content and in particular
thereafter to compute the related correlation functions. Recently we
investigated \cite{CFK,CF1,CF2} a class of models, the $SU(3)_{2}/U(1)^{2}$%
-homogeneous Sine-Gordon model \cite{HSG} (HSG), for which this task was
completed to a large extend. In particular we provided general formulae for
the $n$-particle form factors related to a huge class of local operators. In
order to understand the generic group theoretical structure of the $n$%
-particle form factor expressions it is highly desirable to extend that
analysis to higher rank as well as to higher level. One of the main purposes
of this manuscript is to do the former, that is the investigation of the $%
SU(N)_{2}/U(1)^{N-1}$ case. This model may be viewed as the perturbation of
a gauged WZNW-coset with Virasoro central charge 
\begin{equation}
c_{SU(N)_{2}/U(1)^{N-1}}=\frac{N(N-1)}{(N+2)}
\end{equation}
by an operator of conformal dimension $\Delta =N/(N+2)$. The theory
possesses already a fairly rich particle content, namely $N-1$
asymptotically stable particles characterized by a mass scale $m_{i}$ and $%
N-2$ unstable particles whose energy scale is characterized by the resonance
parameters $\sigma _{ij}$ ($1\leq i,j\leq N-1$). We relate the stable
particles in a one-to-one fashion to the vertices of the $SU(N)$-Dynkin
diagram and associate to the link between vertex $i$ and $j$ the $N-2$
linearly independent resonance parameters $\sigma _{ij}$.

We find that once an unstable particle becomes extremely heavy the original
coset decouples into a direct product of two cosets different from the
original one 
\begin{eqnarray}
&&\lim\limits_{\sigma _{i,i+1}\rightarrow \infty }SU(N)_{2}/U(1)^{N-1}\equiv
\label{flow} \\
&&SU(i+1)_{2}/U(1)^{i}\otimes SU(N-i)_{2}/U(1)^{N-i-1}\,.  \nonumber
\end{eqnarray}
Equivalently we may summarize the flow along the renormalization group
trajectory with increasing RG-parameter $r_{0}$ to cutting the related
Dynkin diagrams at decreasing values of the $\sigma $'s. For instance taking 
$\sigma _{i,i+1}$ to be the largest resonance parameter at some energy scale
the following cut takes place:

\unitlength=1.0cm 
\begin{picture}(8.20,1.0)(-0.5,0.00)
\put(0.00,0.00){\circle*{0.2}}
\put(-0.10,-0.50){$\alpha_1$}
\put(0.00,-0.01){\line(1,0){1.}}
\put(1.00,0.00){\circle*{0.2}}
\put(0.90,-0.50){$\alpha_2$}
\put(0.20,0.30){$\sigma_{12}$}
\put(1.00,-0.01){\line(1,0){0.5}}
\put(1.6,0.00){$\ldots$}
\put(2.70,0.30){$\sigma_{i,i+1}$}
\put(2.2,-0.01){\line(1,0){0.5}}
\put(2.70,0.00){\circle*{0.2}}
\put(2.60,-0.50){$\alpha_i$}
\put(2.70,-0.01){\line(1,0){1.0}}
\put(3.70,0.00){\circle*{0.2}}
\put(3.10,-1.20){$\downarrow$}
\put(3.60,-0.50){$\alpha_{i+1}$}
\put(3.70,-0.01){\line(1,0){0.5}}
\put(4.3,0.00){$\ldots$}
\put(4.90,-0.01){\line(1,0){0.5}}
\put(5.40,0.00){\circle*{0.2}}
\put(5.30,0.30){$\sigma_{N-2,N-1}$}
\put(5.40,-0.01){\line(1,0){1.0}}
\put(6.40,0.00){\circle*{0.2}}
\put(6.30,-0.50){$\alpha_{N-1}$}
\put(5.20,-0.50){$\alpha_{N-2}$}
\end{picture}

\unitlength=1.0cm 
\begin{picture}(8.20,1.5)(-0.5,0.00)
\put(0.00,0.00){\circle*{0.2}}
\put(-0.10,-0.50){$\alpha_1$}
\put(0.00,-0.01){\line(1,0){1.}}
\put(1.00,0.00){\circle*{0.2}}
\put(0.90,-0.50){$\alpha_2$}
\put(1.00,-0.01){\line(1,0){0.5}}
\put(1.6,0.00){$\ldots$}
\put(2.2,-0.01){\line(1,0){0.5}}
\put(2.70,0.00){\circle*{0.2}}
\put(2.60,-0.50){$\alpha_i$}
\put(3.70,0.00){\circle*{0.2}}
\put(3.60,-0.50){$\alpha_{1}$}
\put(3.70,-0.01){\line(1,0){0.5}}
\put(4.3,0.00){$\ldots$}
\put(4.90,-0.01){\line(1,0){0.5}}
\put(5.40,0.00){\circle*{0.2}}
\put(5.40,-0.01){\line(1,0){1.0}}
\put(6.40,0.00){\circle*{0.2}}
\put(6.20,-0.50){$\alpha_{N-i-1}$}
\end{picture}

\vspace*{1.2cm}

Using the usual expressions for the coset central charge \cite{GKO}, the
decoupled system has the central charge {\small 
\begin{equation}
\lim\limits_{\sigma _{i,i+1}\rightarrow \infty
}\!\!\!\!c_{SU(N)_{2}/U(1)^{N-1}}=N-5+\frac{6(N+5)}{(N+2-i)(3+i)}\,.
\end{equation}
} \ 

Our manuscript is organized as follows: In section II we present the main
characteristics of the HSG-scattering matrix. In section III we
systematically construct solutions to the form factor consistency equations,
which correspond to a huge class of local operators. In section IV we
investigate the renormalization group flow of the Virasoro central charge,
reproducing the decoupling (\ref{flow}). In section V we compute the
operator content in terms of primary fields of the underlying conformal
field theory. In section VI we investigate the RG flow of conformal
dimensions. Our conclusions are stated in section VII.

\section{The S-matrix}

The prerequisite for the computation of form factors and correlation
functions thereafter is the knowledge of the exact scattering matrix. The
two-particle S-matrix describing the scattering of two stable particles of
type $i$ and $j$, with $1\leq i,j\leq N-1$, as a function of the rapidity $%
\theta $ related to this model was proposed in \cite{HSGS}. Adopted to a
slightly different notation it may be written as 
\begin{equation}
S_{ij}(\theta )=(-1)^{\delta _{ij}}\left[ c_{i}\tanh \frac{1}{2}\left(
\theta +\sigma _{ij}-i\frac{\pi }{2}\right) \right] ^{I_{ij}}\,.
\label{ZamS}
\end{equation}
The incidence matrix of the $SU(N)$-Dynkin diagram is denoted by $I$. The
parity breaking which is characteristic for the HSG models and manifests
itself by the fact that $S_{ij}\neq S_{ji}$, takes place through the
resonance parameters $\sigma _{ij}=-\sigma _{ji}$ and the colour value $%
c_{i} $. The latter quantity arises from a partition of the Dynkin diagram
into two disjoint sets, which we refer to as ``$+$'' and ``$-$''. We then
associate the values $c_{i}=\pm 1$ to the vertices $i$ of the Dynkin diagram
of $SU(N)$, in such a way that no two vertices related to the same set are
linked together. Likewise we could simply divide the particles into odd and
even, however, such a division would be specific to $SU(N)$ and the
bi-colouration just outlined admits a generalization to other groups as
well. The resonance poles in $S_{ij}(\theta )$ at $(\theta
_{R})_{ij}=-\sigma _{ij}-i\pi /2$ are associated in the usual Breit-Wigner
fashion to the $N-2$ unstable particles as explained for instance in \cite
{ELOP,HSGS,CF2}. It is important for us to recall that the mass of the
unstable particle $M_{\tilde{c}}\,\ $formed in the scattering between the
stable particles $i$ and $j$ behaves as $M_{\tilde{c}}\sim e^{\left| \sigma
_{ij}\right| /2}$. There are no poles present on the imaginary axis, which
indicates that no stable bound states may be formed.

It is clear from the expression of the scattering matrix (\ref{ZamS}), that
whenever a resonance parameter $\sigma _{ij}$ with $I_{ij}\neq 0$ goes to
infinity, we may view the whole system as consisting out of two sets of
particles which only interact freely amongst each other. The unstable
particle, which was created in interaction process between these two
theories before taking the limit, becomes so heavy that it can not be formed
anymore at any energy scale.

\section{Form factors}

We are now in the position to compute the $n$-particle form factors related
to this model, i.e. the matrix elements of a local operator ${\cal O}(\vec{x}%
)$ located at the origin between a multiparticle in-state of particles
(solitons) of species $\mu $, created by $V_{\mu }(\theta )$, and the vacuum

\begin{equation}
F_{n}^{{\cal O}|\mu _{1}\ldots \mu _{n}}(\theta _{1}\ldots \theta
_{n})=\left\langle {\cal O}(0)V_{\mu _{1}}(\theta _{1})V_{\mu _{2}}(\theta
_{2})\ldots V_{\mu _{n}}(\theta _{n})\right\rangle .  \label{ffdef}
\end{equation}

\noindent We proceed in the usual fashion by solving the form factor
consistency equations \cite{Kar,Smir}. For this purpose we extract
explicitly, according to standard procedure, the singularity structure.
Since no stable bound states may be formed during the scattering of two
stable particles the only poles present are the ones associated to the
kinematic residue equations, that is a first order pole for particles of the
same type whose rapidities differ by $i\pi $. Therefore, we parameterize the 
$n$-particle form factors as 
\begin{eqnarray}
F_{n}^{{\cal O}|{\frak {M}}(l_{+},l_{-})}(\theta _{1}\ldots \theta _{n})
&=&H_{n}^{{\cal O}|{\frak {M}}(l_{+},l_{-})}Q_{n}^{{\cal O}|{\frak {M}}%
(l_{+},l_{-})}({\frak {X})}  \label{fact} \\
&&{\frak \,\!\!}\times \prod_{1\leq i<j\leq n}\frac{F_{\text{min}}^{\mu
_{i}\mu _{j}}(\theta _{ij})}{\left( x_{i}^{c_{\mu _{i}}}+x_{j}^{c_{\mu
_{j}}}\right) ^{\delta _{\mu _{i}\mu _{j}}}}\,\,\,.  \nonumber
\end{eqnarray}
As usual we abbreviate the rapidity difference as $\theta _{ij}=\theta
_{i}-\theta _{j}$. Aiming towards a universally applicable and concise
notation, it is convenient to collect the particle species $\mu _{1}\ldots
\mu _{n}$ in form of particular sets 
\begin{eqnarray}
{\frak {M}_{i}}(l_{i}) &=&\{\mu \,|\,\mu =i\} \\
{\frak {M}_{\pm }}(l_{\pm }) &=&\bigcup_{i\in \pm }{\frak {M}_{i}}(l_{i}) \\
{\frak {M}}(l_{+},l_{-}) &=&{\frak {M}_{+}}(l_{+})\cup {\frak {M}_{-}}%
(l_{-}).
\end{eqnarray}
The number of elements belonging to the sets ${\frak {M}_{i},}{\frak M}%
{\frak _{\pm }}$ is indicated by their arguments $l_{i},l_{\pm }$,
respectively. We understand here that inside the sets ${\frak {M}_{\pm }}$
the order of the individual sets ${\frak {M}_{i}}$ is arbitrary. This simply
reflects the fact that particles of different species but identical colour
interact freely. However, ${\frak {M}}$ is an ordered set since elements of $%
{\frak {M}_{+}}$ and ${\frak {M}_{-}}$ do not interact freely and w.l.g. we
adopt the convention that particles belonging to the ``$+$''-colour set come
first. The $H_{n}$ are some overall constants and the $Q_{n}$ are polynomial
functions depending on the variables $x_{i}=\exp \theta _{i}$ which are
collected in the sets ${\frak {X},}{\frak X}{\frak _{i},}{\frak X}{\frak %
_{\pm }}$ in a one-to-one fashion with respect to the particle species sets $%
{\frak {M},}{\frak M}{\frak _{i},}{\frak M}{\frak _{\pm }}$. The functions $%
F_{\text{min}}^{\mu _{i}\mu _{j}}(\theta _{ij})$ are the so-called minimal
form factors which by construction contain no singularities in the physical
sheet and solve Watson's equations \cite{Kar,Smir} for two particles. For
the $SU(N)_{2}$-HSG model they are found to be 
\begin{equation}
F_{\text{min}}^{ij}(\theta )={\cal N}^{I_{ij}}\left( \sin \frac{\theta }{2i}%
\right) ^{\delta _{ij}}e^{-I_{ij}\int\limits_{0}^{\infty }{{\frac{dt}{t}}}%
{\textstyle{\sin ^{2}\left( (i\pi -\theta \mp \sigma _{ij})\frac{t}{2\pi }\right)  \over \sinh t\cosh t/2}}%
}\,.  \label{14}
\end{equation}
Here ${\cal N}=2^{\frac{1}{4}}\exp \left( i\pi (1-c_{i})/4+c_{i}\theta
/4-G/4\right) $ is a normalization function with $G$ being the Catalan
constant. It is convenient also to introduce the function $\tilde{F}_{\text{%
min}}^{ij}(\theta )=(e^{-c_{i}\theta /4}F_{\text{min}}^{ij}(\theta
))^{I_{ij}}$. The minimal form factors obey the functional identities 
{\small 
\begin{equation}
F_{\text{min}}^{ij}(\theta +i\pi )F_{\text{min}}^{ij}(\theta )=\left( -\frac{%
i}{2}\sinh \theta \right) ^{\delta _{ij}}\left( \frac{i^{\frac{2-c_{i}}{2}%
}\,e^{\frac{\theta }{2}c_{i}}}{\cosh \left( \frac{\theta }{2}-\frac{i\,\pi }{%
4}\right) }\right) ^{I_{ij}}.  \label{he2}
\end{equation}
} Substituting the ansatz (\ref{fact}) into the kinematic residue equation 
\cite{Kar,Smir,CFK}, we obtain with the help of (\ref{he2}) a recursive
equation for the overall constants for \thinspace $\mu _{i}\in {\frak {M}_{+}%
}$ 
\begin{equation}
H_{n+2}^{{\cal O}|{\frak {M}(}l_{+}+2,l_{-}{\frak )}}=i^{\bar{l}%
_{i}}2^{2l_{i}-\bar{l}_{i}+1}e^{I_{ij}\sigma _{ij}l_{j}/2}H_{n}^{{\cal O}|%
{\frak {M}(}l_{+},l_{-}{\frak )}}\,\,{\frak \,\,.}  \label{Hrec}
\end{equation}
We introduced here the numbers $\bar{l}_{i}=\sum_{\mu _{j}\in {\frak {M}_{-}}%
}I_{ij}l_{j}$, which count the elements in the neighbouring sets of ${\frak {%
M}_{i}}$.

The $Q$-polynomials have to obey the recursive equations 
\begin{eqnarray}
&&Q_{n+2}^{{\cal O}|{\frak {M}}(2+l_{+},l_{-})}({\frak {X}}%
^{xx})=\sum_{k=0}^{\bar{s}_{i}}x^{2s_{i}-2k+\tau _{i}+1-\varsigma
_{i}}\sigma _{2k+\varsigma _{i}}(I_{ij}{\frak \hat{X}}_{j})  \nonumber \\
&&\quad \qquad \times (-i)^{2s_{i}+\tau _{i}+1}\sigma _{2s_{i}+\tau _{i}}(%
{\frak X}_{i})Q^{{\cal O}|{\frak M}(l_{+},l_{-})}{\frak ({X})\,}  \label{dd}
\end{eqnarray}
For convenience we defined the sets ${\frak {X}}^{xx}:=\{-x,x\}\cup {X}$, $%
{\frak {\hat{X}}:=}ie^{\sigma _{i,i+1}}{\frak X}$ and the integers $\zeta
_{i}$ which are $0$ or $1$ depending on whether the sum $\vartheta +\tau
_{i} $ is odd or even, respectively. $\vartheta $ is related to the factor
of local commutativity $\omega =(-1)^{\vartheta }=\pm 1$. $\sigma
_{k}(x_{1},\ldots ,x_{m})$ is the $k$-th elementary symmetric polynomial.
Furthermore, we used the sum convention $I_{ij}{\frak {\hat{X}}}_{j}\equiv 
\mathop{\textstyle\bigcup}%
_{\mu _{j}\in {\frak M}}I_{ij}{\frak \hat{X}}_{j}$ and parameterized $%
l_{i}=2s_{i}+\tau _{i}$, $\bar{l}_{i}=2\bar{s}_{i}+\bar{\tau}_{i}$ in order
to distinguish between odd and even particle numbers.

We will now solve the recursive equations (\ref{Hrec}) and (\ref{dd})
systematically. The equations for the constants are solved by 
\begin{equation}
H_{n}^{{\cal O}|{\frak {M}(}l_{+},l_{-}{\frak )}}=\!\!\!\!\prod\limits_{\mu
_{i}\in {\frak M}_{+}}\!\!\!\!\!\!i^{s_{i}\bar{l}_{i}}\,2^{s_{i}(2s_{i}-\bar{%
l}_{i}-1-2\tau _{i})}e^{\frac{s_{i}I_{ij}\sigma _{ij}l_{j}}{2}}H^{{\cal O}|{%
\tau }_{i},\bar{l}_{i}}.
\end{equation}
The lowest nonvanishing constants $H^{{\cal O}|{\tau }_{i},\bar{l}_{i}}$ are
fixed by demanding, similarly as in the $SU(3)_{2}$-case \cite{CFK,CF1,CF2},
that any form factor which involves only one particle type should correspond
to a form factor of the thermally perturbed Ising model. To achieve this we
exploit the ambiguity present in (\ref{Hrec}), that is the fact that we can
multiply it by any constant which only depends on the $l_{-}$-quantum
numbers.

As the main building blocks for the construction of the $Q$-polynomials
serve the determinants of the ($t+s$)$\times $($t+s$)-matrix {\small 
\begin{equation}
{\cal \,A}_{2s+\tau ^{+},2t+\tau ^{-}}^{\nu ^{+},\nu ^{-}}({\frak {X}}_{+},%
{\frak {\hat{X}}}_{-})_{ij}=%
{\sigma _{2(j-i)+\nu ^{+}}({\frak {X}}_{+})\text{,}1\leq i\leq t \atopwithdelims\{. \sigma _{2(j-i+t)+\nu ^{-}}({\frak {\hat{X}}}_{-})\,\text{, }t<i\leq s+t}%
\end{equation}
}for $\nu ^{\pm },\tau ^{\pm }=0,1$ which were introduced in \cite{CF1}. The
determinant of ${\cal A}$ essentially captures the summation in (\ref{dd})
due to the fact that it satisfies the recursive equations 
\begin{eqnarray}
&&\det {\cal A}_{2+l,2t+\tau ^{-}}^{\nu ^{+},\nu ^{-}}({\frak {X}}%
_{+}^{\,xx},{\frak \hat{X}}_{-})\,\,=\left(
\sum\limits_{p=0}^{t}x^{2(t-p)}\sigma _{2p+\nu ^{-}}({\frak \hat{X}}%
_{-})\right) \,  \nonumber \\
&&\qquad \quad \qquad \qquad \qquad \qquad \times \det {\cal A}_{l,2t+\tau
^{-}}^{\nu ^{+},\nu ^{-}}({\frak X}_{+},{\frak \hat{X}}_{-})\,  \label{xx}
\end{eqnarray}
as was shown in \cite{CF1}. Analogously to the procedure in \cite{CF1} we
can build up a simple product from elementary symmetric polynomials which
takes care of the pre-factor in the recursive equation (\ref{dd}). Defining
therefore the polynomials 
\begin{eqnarray}
&&Q_{n}^{{\frak {M}}(l_{+},l_{-})}({\frak {X}}_{+},{\frak X}%
_{-})=\!\!\!\!\prod\limits_{\mu _{i/k}\in {\frak M}_{+/-}}\!\!\!\!\!\!\!\!\!%
\det {\cal A}_{2s_{i}+\tau _{i},\bar{l}_{i}}^{\nu _{i},\varsigma _{i}}(%
{\frak X}_{i},I_{ij}{\frak \hat{X}}_{j})  \label{sol} \\
&&\times \sigma _{2s_{i}+\tau _{i}}({\frak X}_{i})^{\frac{2s_{i}+\tau _{i}-2%
\bar{s}_{i}-1-\varsigma _{i}}{2}}\sigma _{\bar{l}_{i}}(I_{ij}{\frak \hat{X}}%
_{j})^{\frac{\bar{\nu}_{i}-1}{2}}\sigma _{l_{k}}({\frak \hat{X}}_{k}))^{%
\frac{1-l_{k}}{2}}  \nonumber
\end{eqnarray}
it follows immediately with the help of property (\ref{xx}) that they obey
the recursive equations 
\begin{eqnarray}
&&Q_{2+n}^{{\frak {M}}(2+l_{+},l_{-})}({\frak {X}}_{+}^{xx},{\frak X}%
_{-})=Q_{n}^{{\frak {M}}(l_{+},l_{-})}({\frak X}_{+},{\frak X}_{-})\sigma
_{2s_{i}+\tau _{i}}({\frak X}_{i})\,  \nonumber \\
&&\qquad \qquad \qquad \times \sum\limits_{p=0}^{\bar{s}_{i}}x^{2(s_{i}-p)+%
\tau _{i}+1-\varsigma _{i}}\sigma _{2p+\varsigma _{i}}(I_{ij}{\frak \hat{X}}%
_{j})\,\,.  \label{an}
\end{eqnarray}
Comparing now the equations (\ref{dd}) and (\ref{an}) we obtain complete
agreement. Notice that the numbers $\bar{\nu}_{i}$ are not constrained at
all at this point of the construction. However, by demanding relativistic
invariance, which on the other hand means that the overall power in (\ref
{fact}) has to be zero, we obtain the additional constraints 
\begin{equation}
\nu _{i}=1+\tau _{i}-\bar{\nu}_{i}\qquad \text{and\qquad }\tau _{i}\varsigma
_{i}=\bar{\tau}_{i}(\bar{\nu}_{i}-1)\,\,\,.
\end{equation}
Taking in addition the constraints into account which are needed to derive (%
\ref{xx}) (see \cite{CF1}), this is most conveniently written as 
\begin{equation}
\tau _{i}\varsigma _{i}+\bar{\tau}_{i}\nu _{i}=\tau _{i}\bar{\tau}%
_{i},\qquad 2+\varsigma _{i}>\bar{\tau}_{i},\qquad 2+\nu _{i}>\tau
_{i}\,\,\,.  \label{consist}
\end{equation}
For each $\mu _{i}\in {\frak M}_{+}$ the equations (\ref{consist}) admit the
10 feasible solutions found in \cite{CF1}. However, one should notice that
the individual solutions for different values of $i$ are not all independent
of each other. We would like to stress that despite the fact that (\ref{sol}%
) represents a huge class of independent solutions, it does certainly not
exhaust all of them. Nonetheless, many additional solutions, like the energy
momentum tensor, may be constructed from (\ref{consist}) by simple
manipulations like the multiplication of some CDD-like ambiguity factors or
by setting some expressions to zero on the base of asymptotic considerations
(see \cite{CF1}) for more details. For many applications we wish to carry
out in the next section we require the form factors for the trace of the
energy momentum $\Theta $, the first non-vanishing terms read

\begin{eqnarray*}
F_{2}^{\Theta |\mu _{i}\mu _{i}} &=&-2\pi im^{2}\sinh (\theta /2) \\
F_{4}^{\Theta |\mu _{i}\mu _{i}\mu _{j}\mu _{j}} &=&\frac{\pi
m^{2}(2+\sum\limits_{i<j}\cosh (\theta _{ij}))\prod\limits_{i<j}\tilde{F}_{%
\text{min}}^{\mu _{i}\mu _{j}}(\theta _{ij})}{-2\cosh (\theta _{12}/2)\cosh
(\theta _{34}/2)}\, \\
F_{6}^{\Theta |\mu _{i}\mu _{i}\mu _{i}\mu _{i}\mu _{j}\mu _{j}} &=&\frac{%
\pi m^{2}(3+\sum\limits_{i<j}\cosh (\theta _{ij}))\prod\limits_{i<j}\tilde{F}%
_{\text{min}}^{\mu _{i}\mu _{j}}(\theta _{ij})}{4\prod_{1\leq i<j\leq
4}\cosh (\theta _{ij}/2)} \\
F_{6}^{\Theta |\mu _{i}\mu _{i}\mu _{k}\mu _{k}\mu _{j}\mu _{j}} &=&\frac{%
\pi m^{2}(3+\sum\limits_{i<j}\cosh (\theta _{ij}))\prod\limits_{i<j}\tilde{F}%
_{\text{min}}^{\mu _{i}\mu _{j}}(\theta _{ij})}{4\cosh (\theta _{12}/2)\cosh
(\theta _{34}/2)}
\end{eqnarray*}
for $I_{ij}\neq 0$ and $I_{kj}\neq 0$ . When considering the RG flow in the
next section, it will be important to note that from $\lim_{\sigma
_{i,i+1}\rightarrow \infty }F_{\text{min}}^{\mu _{i}\mu _{i+1}}(\theta )\sim
\exp (-\sigma _{i,i+1}/4)$ follows 
\begin{equation}
\lim_{\sigma _{i,i+1}\rightarrow \infty }F_{n}^{\Theta |\mu _{i}\mu
_{i+1}\ldots }(\theta )=0\,\,.  \label{limF}
\end{equation}

Having determined the form factors, we are in principle in the position to
compute the two-point correlation function between two local operators in
the usual way by expanding it in terms of $n$-particle form factors 
\begin{eqnarray}
&&\left\langle {\cal O}(r){\cal O}^{\prime }(0)\right\rangle
=\sum_{n=1}^{\infty }\sum_{\mu _{1}\ldots \mu _{n}}\int\limits_{-\infty
}^{\infty }\frac{d\theta _{1}\ldots d\theta _{n}}{n!(2\pi )^{n}}e^{-r\,E}
\label{corre} \\
&&\times \,\,F_{n}^{{\cal O}|\mu _{1}\ldots \mu _{n}}(\theta _{1},\ldots
,\theta _{n})\,\left( F_{n}^{{\cal O}^{\prime }|\mu _{1}\ldots \mu
_{n}}(\theta _{1},\ldots ,\theta _{n})\,\right) ^{\ast }.  \nonumber
\end{eqnarray}
We abbreviated the sum of the on-shell energies as $E=\sum%
\nolimits_{i=1}^{n}m_{\mu _{i}}\cosh \theta _{i}$. Now we want to evaluate
the expression (\ref{corre}) in several different applications in order to
compute various quantities of interest.

\section{Renormalization group flow}

Renormalization group methods have been developed originally \cite{GL} to
carry out qualitative analysis of regions of quantum field theories which
are not accessible by perturbation theory in the coupling constant. In
particular the $\beta $-function provides an inside into various possible
asymptotic behaviours and especially it allows to identify the fixed points
of the theory. We now want to employ this method to check our solutions and
at the same time the physical picture advocated for the HSG-models.

For this purpose we want to investigate first of all the renormalization
group flow, in a similar spirit as for the $SU(3)_{2}/U(1)^{2}$-case in \cite
{CF2}, by evaluating the c-theorem \cite{ZamC}. 
\begin{equation}
c(r_{0})=\frac{3}{2}\int\limits_{r_{0}}^{\infty }dr\,r^{3}\,\,\left\langle
\Theta (r)\Theta (0)\right\rangle \,\,.  \label{cth}
\end{equation}
In particular for $r_{0}=0$ the function $c(r_{0})$ coincides with $\Delta
c=c_{\text{uv}}-c_{\text{ir}}$, i.e. the difference between the ultraviolet
and infrared Virasoro central charges. Computing the correlation function
for the trace of the energy-momentum tensor $\Theta $ in (\ref{cth}) by
means of (\ref{corre}) and using the form factor expressions of the previous
section the individual $n$-particle contributions turn out to be 
\begin{eqnarray}
\Delta c^{(2)} &=&(N-1)\cdot 0.5  \label{2p} \\
\Delta c^{(4)} &=&(N-2)\cdot 0.197  \label{4p} \\
\Delta c^{(6)} &=&(N-2)\cdot 0.002+(N-3)\cdot 0.0924  \label{6p} \\
\sum_{k=2}^{6}\Delta c^{(k)} &=&N\ast 0.7914-1.1752\,\,.  \label{sum}
\end{eqnarray}
Apart from the two particle contribution (\ref{2p}), which is usually quite
trivial and in this situation can even be evaluated analytically, we have
carried out the multidimensional integrals in (\ref{corre}) by means of a
Monte Carlo method. We use this method up to a precision which is higher
than the last digit we quote. For convenience we report some explicit
numbers in table 1.

\begin{center}
\noindent 
\begin{tabular}{|c||c|c|c|c|c|}
\hline
$N$ & $c$ & $\Delta c^{(2)}$ & $\Delta c^{(4)}$ & $\Delta c^{(6)}$ & $%
\sum_{k=2}^{6}\Delta c^{(k)}$ \\ \hline\hline
$3$ & $1.2$ & $1$ & $0.197$ & $0.002$ & $1.199$ \\ \hline
$4$ & $2$ & $1.5$ & $0.394$ & $0.096$ & $1.990$ \\ \hline
$5$ & $2.857$ & $2$ & $0.591$ & $0.191$ & $2.782$ \\ \hline
$6$ & $3.75$ & $2.5$ & $0.788$ & $0.285$ & $3.573$ \\ \hline
$7$ & $4.\bar{6}$ & $3$ & $0.985$ & $0.380$ & $4.365$ \\ \hline
$8$ & $5.6$ & $3.5$ & $1.182$ & $0.474$ & $5.156$ \\ \hline
\end{tabular}
\end{center}

\noindent {\small Table 1: n-particle contributions to the c-theorem versus $%
SU(N)_{2}/U(1)^{N-1}$ -WZNW coset model central charge.}

\medskip

The evaluation of (\ref{2p})-(\ref{sum}) illustrates that the series (\ref
{corre}) converges slower and slower for increasing values of $N$, such that
the higher $n$-particle contributions become more and more important to
achieve high accuracy. Our analysis suggests that it is not the functional
dependence of the individual form factors which is responsible for this
behaviour. Instead this effect is simply due to the fact that the symmetry
factor, that is the sum $\sum_{\mu _{1}\ldots \mu _{n}}$, resulting from
permutations of the particle species increases drastically for larger $N$.

Having confirmed the expected ultraviolet central charge, we now study the
RG-flow by varying $r_{0}$ in (\ref{cth}). We expect to find that whenever
we reach an energy scale at which an unstable particle can be formed, the
model will flow to a different coset. This means following the flow with
increasing $r_{0}$ we will encounter a situation in which certain $\sigma
_{i,i+1}$ are considered to be large and we observe the decoupling into two
freely interacting systems in the way described in (\ref{flow}). For
instance for the situation $\sigma _{12}>\sigma _{23}>\sigma _{34}>\ldots $
we observe the following decoupling along the flow with increasing $r_{0}$:

\begin{center}
\begin{tabular}{c}
$SU(N)_{2}/U(1)^{N-1}$ \\ 
$\downarrow $ \\ 
$SU(N-1)_{2}/U(1)^{N-2}\otimes SU(2)_{2}/U(1)$ \\ 
$\downarrow $ \\ 
$SU(N-2)_{2}/U(1)^{N-3}\otimes \left( SU(2)_{2}/U(1)\right) ^{2}$ \\ 
$\downarrow $ \\ 
$\vdots $ \\ 
$\downarrow $ \\ 
$\left( SU(2)_{2}/U(1)\right) ^{N-1}$%
\end{tabular}
\end{center}

We can understand this type of behaviour in a semi-analytical way. The
precise difference between the central charges related to (\ref{flow}) is 
\begin{eqnarray}
&&c_{SU(i+1)_{2}/U(1)^{i}\otimes SU(N-i)_{2}/U(1)^{N-i-1}}=  \label{diffcex}
\\
&&\qquad c_{SU(N)_{2}/U(1)^{N-1}}-\frac{2i(N+5)(N-i-1)}{(N+2)(i+3)(N-i+2)}. 
\nonumber
\end{eqnarray}
Noting with (\ref{limF}) that we loose at each step all the contributions $%
F_{n}^{\Theta |\mu _{i}\mu _{i+1}\ldots }(\theta )$ to $\Delta c$, we may
collect the values (\ref{2p})-(\ref{6p}), which we have determined
numerically and find 
\begin{eqnarray}
&&\lim_{\sigma _{i,i+1}\rightarrow \infty }{\small \Delta c(\sigma }_{i,i+1}%
{\small ,\ldots )}=  \label{diffc} \\
&&{\small \Delta c(\sigma }_{i,i+1}={\small 0,\ldots )-0.2914I}_{i,i+1}%
{\small -0.0924I}_{i,i-1}  \nonumber
\end{eqnarray}

Similarly as for the deep UV-region we find a relatively good agreement
between (\ref{diffcex}) and (\ref{diffc}) for small values of $N$. The
difference for larger values is once again due to the convergence behaviour
of the series in (\ref{corre}).

For $r_{0}=0$ qualitatively a similar kind of behaviour was previously
observed, for the two particle contribution only, in the context of the
roaming Sinh-Gordon model \cite{ADM}. Nonetheless, there is a slight
difference between the two situations. Instead of a decoupling into
different cosets in these type of models the entire S-matrix takes on the
value $-1$, when the resonance parameter goes to infinity. The resulting
effect, i.e. a depletion of $\Delta c$, is the same. However, we do not
comply with the interpretation put forward in \cite{ADM}, namely that such a
behaviour should constitute a ``violation of the c-theorem''. The observed
effect is precisely what one expects from the physical point of view and the
c-theorem.

We present our full numerical results in figure 1, which confirm the
outlined flow for various values of N.

\begin{center}
\includegraphics[width=8.2cm,height=6.09cm]{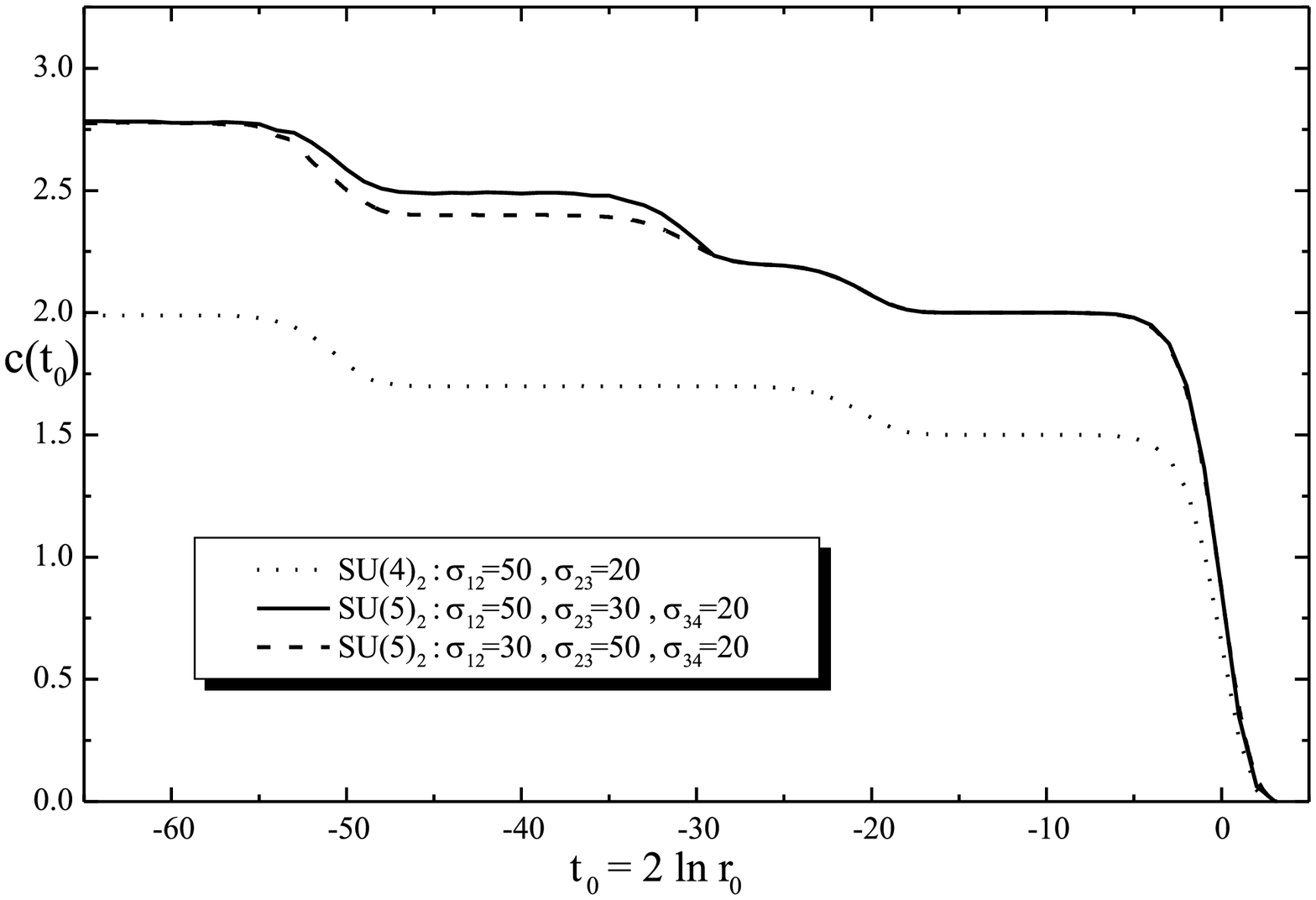}
\end{center}

\vspace*{0.2cm}

\noindent {\small Figure 1: RG flow for the Virasoro central charge.}

\vspace*{1.2mm}

We observe that the $c$-function remains constant, at a value corresponding
to the new coset, in some finite interval of $r_{0}$. In particular, we
observe the non-equivalence of the flows when the relative order of
magnitude amongst the different resonance parameters is changed. For $N=5$
we confirm (we omit here the $U(1)$-factors and report the corresponding
central charges as superscripts on the last factor)

\begin{center}
\begin{tabular}{c}
\frame{$\sigma _{12}>\sigma _{23}$}$\quad SU(5)_{2}^{\frac{20}{7}}\quad 
\frame{$\sigma _{23}>\sigma _{12}$}$ \\ 
$\swarrow \qquad \qquad \searrow $ \\ 
$\quad \quad SU(4)_{2}\otimes SU(2)_{2}^{\frac{5}{2}}\quad \quad \quad \quad
\quad \qquad SU(3)_{2}\otimes SU(3)_{2}^{\frac{12}{5}}\qquad $ \\ 
$\searrow \qquad \qquad \qquad \swarrow $ \\ 
$SU(3)_{2}\otimes SU(2)_{2}\otimes SU(2)_{2}^{\frac{11}{5}}$ \\ 
$\downarrow $ \\ 
$SU(2)_{2}\otimes SU(2)_{2}\otimes SU(2)_{2}\otimes SU(2)_{2}^{2}$%
\end{tabular}
\end{center}

The precise difference in the central charges is explained with (\ref{diffc}%
), since the contribution $0.0924I_{i,i-1}$ only occurs for $i=1$.

To establish more clearly that the plateaus admit indeed an interpretation
as fixed points and extract the definite values of the corresponding
Virasoro central charge we can also, following \cite{ZamC}, determine a $%
\beta $ type function from $c(r)$. The $\beta $-function should obey the
Callan-Symanzik equation \cite{CS} 
\begin{equation}
r\frac{d}{dr}g=\beta (g)\,\,.  \label{CS}
\end{equation}

$\,\,$The ``coupling constant'' $g:=c_{\text{uv}}-c(r)$ is normalized in
such a way that it vanishes at the ultraviolet fixed point. Whenever we find 
$\beta (\tilde{g})=0$, we can identify $\tilde{c}=c_{\text{uv}}-\tilde{g}$
as the Virasoro central charge of the corresponding conformal field theory.
Hence, taking the data obtained from (\ref{cth}), we compute $\beta $ as a
function of $g$ by means of (\ref{CS}). Our results for various values of N
are depicted in figure 2, which allow a definite identification of the fixed
points corresponding to the coset models expected from the decoupling (\ref
{flow}).

\begin{center}
\includegraphics[width=8.2cm,height=6.09cm]{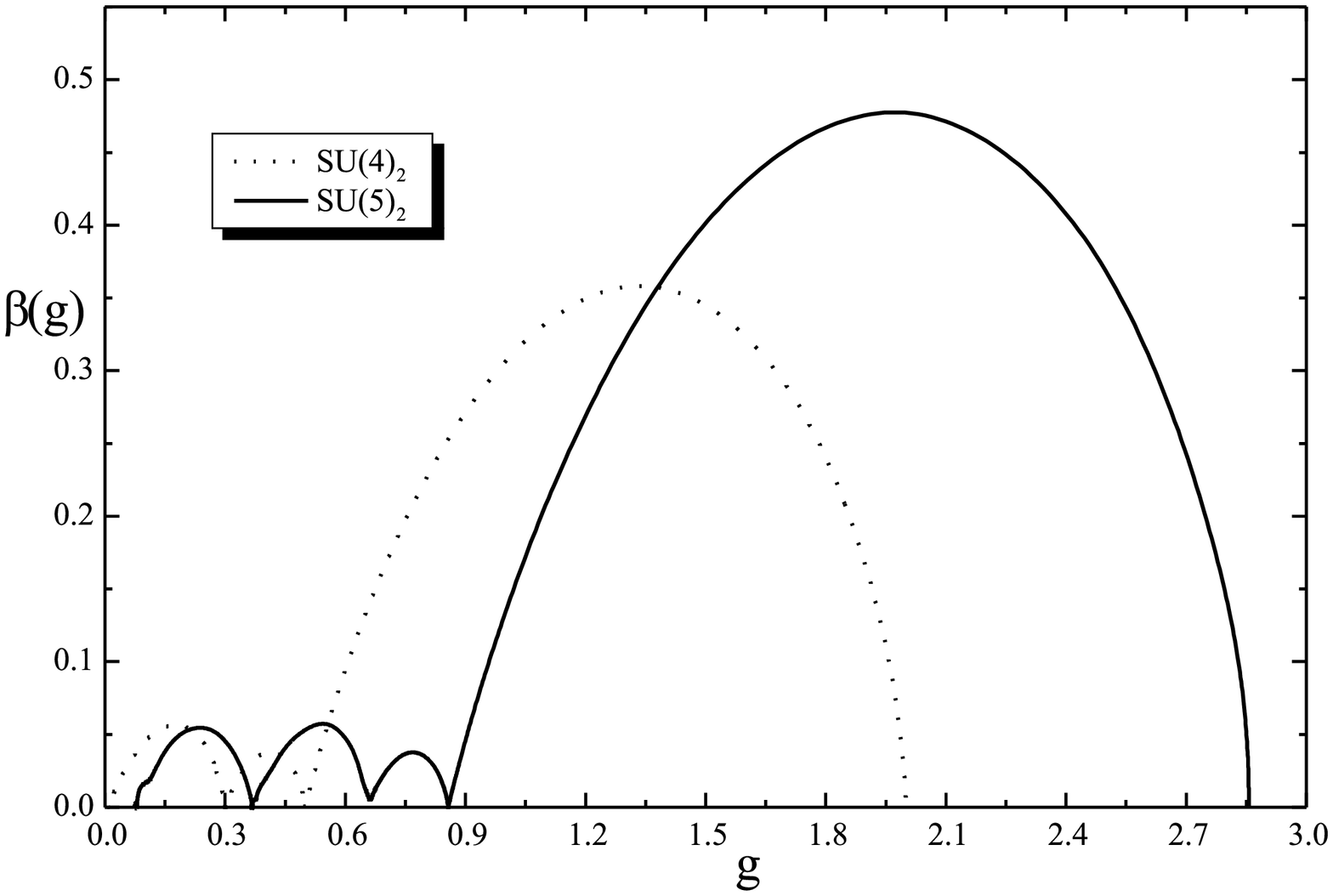}
\end{center}

\vspace*{0.2cm}

\noindent {\small Figure 2: The $\beta $-function.}

For $SU(4)_{2}$ we clearly identify from figure 2 the four fixed points 
{\small \ }$\tilde{g}=0,0.3,0.5,2$ with high accuracy. The five fixed points 
$\tilde{g}=0,0.357,0.657,0.857,2.857$, which we expect to find for $%
SU(5)_{2} $ are all slightly shifted due to the absence of the higher order
contributions.

\section{Operator content of $SU(N)_{2}/U(1)^{N-1}$}

We now want to identify the operator content of our theory by carrying out
the ultraviolet limit and matching the conformal dimension of each operator
with the one in the $SU(N)_{2}/U(1)^{N-1}$-WZNW-coset model. For this
purpose we have to determine first of all the entire operator content of the
conformal field theory.

According to \cite{Gep} the conformal dimensions of the parafermionic vertex
operators are given by 
\begin{equation}
\Delta (\Lambda ,\lambda )=\frac{(\Lambda \cdot (\Lambda +2\rho ))}{(4+2N)}-%
\frac{(\lambda \cdot \lambda )}{4}\,\,.  \label{dim}
\end{equation}
Here $\Lambda $ is a highest dominant weight of level smaller or equal to $2$
and $\rho =1/2\sum_{\alpha >0}\alpha $ is the Weyl vector, i.e. half the sum
of all positive roots. The $\lambda $'s are the corresponding lower weights,
which may be constructed in the usual fashion (see e.g. \cite{FH}): Consider
a complete weight string $\lambda +n\alpha ,\ldots ,\lambda ,\ldots ,\lambda
-m\alpha $, that is all the weights obtained by successive additions
(subtractions) of a root $\alpha $ from the weight $\lambda $, such that $%
\lambda +(n+1)\alpha $ ($\lambda -(m+1)\alpha $) is not a weight anymore. It
is then a well known fact that the difference between the two integers $m,n$
is $m-n=\lambda \cdot \alpha $ for simply laced Lie algebras. This means
starting with the highest weight $\Lambda $, we can work our way downwards
by deciding after each subtraction of a simple root $\alpha _{i}$ whether
the new vector, say $\chi $, is a weight or not from the criterion $%
m_{i}=n_{i}+\chi \cdot \alpha _{i}>0$. With the procedure just outlined we
obtain all possible weights of the theory. Nonetheless, it may happen that a
weight corresponds to more than one linear independent weight vector, such
that the weight space may be more than one dimensional. The dimension of
each weight vector $n_{\lambda }^{\Lambda }$ is computed by means of 
\begin{equation}
n_{\lambda }^{\Lambda }=\frac{\sum_{\alpha >0}\sum_{l=1}^{\infty
}2\,n_{\lambda +l\alpha }((\lambda +l\alpha )\cdot \alpha )}{((\Lambda
+\lambda +2\rho )\cdot (\Lambda -\lambda ))}\,\,\,.  \label{mult}
\end{equation}
For consistency it is useful to compare the sum of all these multiplicities
with the dimension of the highest weight representation computed directly
from the Weyl dimensionality formula (see e.g. \cite{FH}) 
\begin{equation}
\sum_{\lambda }n_{\lambda }^{\Lambda }=\dim \Lambda =\prod_{\alpha >0}\frac{%
((\Lambda +\rho )\cdot \alpha )}{(\rho \cdot \alpha )}\,\,.  \label{weyl}
\end{equation}
To compute all the conformal dimensions $\Delta (\Lambda ,\lambda )$
according to (\ref{dim}) in general is a formidable task and therefore we
concentrate on a few distinct ones for generic $N$ and only compute the
entire content for $N=4$.

Noting that $\,\lambda _{i}\cdot \,\lambda _{j}=K_{ij}^{-1}$, with $K$ being
the Cartan matrix, we can obtain relative concrete formulae from (\ref{dim}%
). For instance 
\begin{equation}
\Delta (\lambda _{i},\lambda _{i})=\frac{%
4\sum_{l=1}^{N-1}K_{il}^{-1}-NK_{ii}^{-1}}{8+2N}\,\,\,.
\end{equation}

Similarly we may compute $\Delta (\lambda _{i}+\lambda _{j},\lambda
_{i}+\lambda _{j})$, etc. in terms of components of the inverse Cartan
matrix. Even more explicit formulae are obtainable when we express the
simple roots $\alpha _{i}$ and fundamental weights $\lambda _{i}$ of $SU(N)$
in terms of a concrete basis. For instance we may choose an orthonormal
basis \{$\varepsilon _{i}$\} in ${\Bbb R}^{N}$ (see e.g. \cite{Bou}), i.e. $%
\varepsilon _{i}\cdot \varepsilon _{j}=\delta _{ij}$ 
\[
\alpha _{i}=\varepsilon _{i}-\varepsilon _{i+1},\,\,\,\lambda
_{i}=\sum\limits_{j=1}^{i}\varepsilon _{j}-\frac{i}{N}\sum\limits_{j=1}^{N}%
\varepsilon _{j},\,\,i=1,\ldots N-1\,. 
\]
Noting further that the set of positive roots is given by $\{\varepsilon
_{i}-\varepsilon _{j}:1\leq i<j\leq N\}$, we can evaluate (\ref{dim}), (\ref
{mult}) and (\ref{weyl}) explicitly. This way we obtain for instance 
\begin{equation}
\Delta (\lambda _{i},\lambda _{i})=\frac{i(N-i)}{8+4N}\,\quad \,\text{%
and\quad }\,\,\Delta (2\lambda _{i},2\lambda _{i})=0.
\end{equation}
Of special physical interest is the dimension of the perturbing operator. As
was already argued in \cite{CF2}, it corresponds to $\Delta (\psi ,0)$, with 
$\psi $ being the highest root, and moreover it is unique. Noting that for $%
SU(N)$ we have $\psi =\lambda _{1}+\lambda _{N-1}$, we confirm once more 
\begin{equation}
\Delta (\psi =\lambda _{1}+\lambda _{N-1},0)=\frac{N}{N+2}\,\,.
\end{equation}
Other dimensions may be computed similarly.

\subsection{ The $SU(4)/U(1)^{3}$ example}

For $SU(4)/U(1)^{3}$ we present the result of the computation of the entire
operator content in table 2. In case the multiplicity of a weight vector is
bigger than one, we indicate this by a superscript on the conformal
dimension.

\begin{center}
\noindent {\small 
\begin{tabular}{|c||c|c|c|c|c|c|}
\hline
$\lambda \backslash \Lambda $ & $\lambda _{1}$ & $\lambda _{2}$ & $2\lambda
_{1}$ & $2\lambda _{2}$ & $\lambda _{1}+\lambda _{2}$ & $\lambda
_{1}+\lambda _{3}$ \\ \hline\hline
dim $\Lambda $ & $4$ & $6$ & $10$ & $20$ & $20$ & $15$ \\ \hline
$\Lambda $ & $1/8$ & $1/6$ & $0$ & $0$ & $1/8$ & $1/6$ \\ \hline
$\Lambda -\alpha _{1}$ & $1/8$ &  & $1/2$ &  & $1/8$ & $1/6$ \\ \hline
$\Lambda -\alpha _{2}$ &  & $1/6$ &  & $1/2$ & $1/8$ &  \\ \hline
$\Lambda -\alpha _{3}$ &  &  &  &  &  & $1/6$ \\ \hline
$\Lambda -\alpha _{1}-\alpha _{2}$ & $1/8$ & $1/6$ & $1/2$ & $1/2$ & $%
5/8^{2} $ & $1/6$ \\ \hline
$\Lambda -\alpha _{2}-\alpha _{3}$ &  & $1/6$ &  & $1/2$ & $1/8$ & $1/6$ \\ 
\hline
$\Lambda -\alpha _{1}-\alpha _{3}$ &  &  &  &  &  & $1/6$ \\ \hline
$\Lambda -2\alpha _{1}$ &  &  & $0$ &  &  &  \\ \hline
$\Lambda -2\alpha _{2}$ &  &  &  & $0$ &  &  \\ \hline
$\Lambda -2\alpha _{1}-2\alpha _{2}$ &  &  & $0$ & $0$ &  &  \\ \hline
$\Lambda -2\alpha _{2}-2\alpha _{3}$ &  &  &  & $0$ &  &  \\ \hline
$\Lambda -2\alpha _{1}-\alpha _{2}$ &  &  & $1/2$ &  & $1/8$ &  \\ \hline
$\Lambda -\alpha _{1}-2\alpha _{2}$ &  &  &  & $1/2$ & $1/8$ &  \\ \hline
$\Lambda -2\alpha _{2}-\alpha _{3}$ &  &  &  & $1/2$ &  &  \\ \hline
$\Lambda -\alpha _{1}-\alpha _{2}-\alpha _{3}$ & $1/8$ & $1/6$ & $1/2$ & $%
1/2 $ & $5/8^{2}$ & $2/3^{3}$ \\ \hline
$\Lambda -2\alpha _{1}-\alpha _{2}-\alpha _{3}$ &  &  & $1/2$ &  & $1/8$ & $%
1/6$ \\ \hline
$\Lambda -\alpha _{1}-2\alpha _{2}-\alpha _{3}$ &  & $1/6$ &  & 1$^{2}$ & $%
5/8^{2}$ & $1/6$ \\ \hline
$\Lambda -\alpha _{1}-\alpha _{2}-2\alpha _{3}$ &  &  &  &  &  & $1/6$ \\ 
\hline
$\Lambda -2\alpha _{1}-2\alpha _{2}-\alpha _{3}$ &  &  & $1/2$ & $1/2$ & $%
5/8^{2}$ & $1/6$ \\ \hline
$\Lambda -\alpha _{1}-2\alpha _{2}-2\alpha _{3}$ &  &  &  & $1/2$ & $1/8$ & $%
1/6$ \\ \hline
$\Lambda -2\alpha _{1}-2\alpha _{2}$ &  &  &  &  & $1/8$ &  \\ \hline
$\Lambda -2\alpha _{1}-2\alpha _{2}-2\alpha _{3}$ &  &  & $0$ & 0 & $1/8$ & $%
1/6$ \\ \hline
$\Lambda -\alpha _{1}-3\alpha _{2}-2\alpha _{3}$ &  &  &  & $1/2$ &  &  \\ 
\hline
$\Lambda -\alpha _{1}-3\alpha _{2}-\alpha _{3}$ &  &  &  & $1/2$ &  &  \\ 
\hline
$\Lambda -2\alpha _{1}-3\alpha _{2}-\alpha _{3}$ &  &  &  & $1/2$ & $1/8$ & 
\\ \hline
$\Lambda -2\alpha _{1}-3\alpha _{2}-2\alpha _{3}$ &  &  &  & $1/2$ & $1/8$ & 
\\ \hline
$\Lambda -2\alpha _{1}-4\alpha _{2}-2\alpha _{3}$ &  &  &  & $0$ &  &  \\ 
\hline
\end{tabular}
}
\end{center}

\noindent {\small Table 2: Conformal dimensions for ${\cal O}^{\Delta
(\Lambda ,\lambda )}$ in the $SU(4)_{2}/U(1)^{3}$ -WZNW coset model.\medskip 
}

The remaining dominant weights of level smaller or equal to $2$, namely $%
\Lambda =\lambda _{3},2\lambda _{3},\lambda _{2}+\lambda _{3}$, including
their multiplicities may be obtained from table 2 simply by the exchange $%
1\leftrightarrow 3$, which corresponds to the ${\Bbb Z}_{2}$-symmetry of the 
$SU(4)$-Dynkin diagram.

Summing up all the fields corresponding to different lower weights, i.e. not
counting the multiplicities, we have the following operator content

{\small 
\[
{\cal O}^{2/3},{\cal O}^{1},14\times {\cal O}^{0},8\times {\cal O}%
^{5/8},18\times {\cal O}^{1/6},24\times {\cal O}^{1/2},32\times {\cal O}%
^{1/8}, 
\]
}that is 98 fields.

\section{Operator Content of HSG}

We will now turn to the massive model and evaluate the flow of the conformal
dimension \cite{CF2} 
\begin{equation}
\Delta ^{{\cal O}}(r_{0})=-\frac{1}{2\left\langle {\cal O}(0)\right\rangle }%
\int\limits_{r_{0}}^{\infty }dr\,r\,\,\left\langle \Theta (r){\cal O}%
(0)\right\rangle \,\,.  \label{dr}
\end{equation}
Here ${\cal O}$ is a local operator which in the conformal limit corresponds
to a primary field in the sense of \cite{BPZ}. In particular for $r_{0}=0$,
the expression (\ref{dr}) constitutes the delta sum rule \cite{DSC}, which
expresses the difference between the ultraviolet and infrared conformal
dimension of the operator ${\cal O}$.

We start by investigating the operator which in the case when all particles
in (\ref{ffdef}) are of the same type corresponds to the disorder operator $%
\mu $ in the Ising model. Using the fact that we should always be able to
reduce to that situation, we consider the solution corresponding to $\tau
_{i}=\bar{\tau}_{i}=\nu _{i}=\varsigma _{i}=0$ for all $i$. Then the $\Delta 
$-sum rule (\ref{dr}) yields for the individual $n$-particle contributions 
\begin{eqnarray}
\Delta ^{\mu (2)} &=&(N-1)\cdot 0.0625  \label{l1} \\
\Delta ^{\mu (4)} &=&(2-N)\cdot 0.0263  \label{l2} \\
\Delta ^{\mu (6)} &=&(N-2)\cdot 0.0017+(3-N)\cdot 0.0113  \label{l3} \\
\sum_{k=2}^{6}\Delta ^{\mu (k)} &=&0.0266+N\ast 0.0206\,\,.  \label{l4}
\end{eqnarray}

We assume that this solution has the conformal dimension $\Delta (\lambda
_{1},\lambda _{1})$ in the ultraviolet limit. For comparison we report a few
explicit numbers in table 3.

\begin{center}
\noindent 
\begin{tabular}{|c||c|c|c|c|c|}
\hline
$N$ & $\Delta (\lambda _{1},\lambda _{1})$ & $\Delta ^{\mu (2)}$ & $\Delta
^{\mu (4)}$ & $\Delta ^{\mu (6)}$ & $\sum_{k=2}^{6}\Delta ^{\mu }$ \\ 
\hline\hline
$3$ & $0.1$ & $0.125$ & $-0.0263$ & $0.0017$ & $0.1004$ \\ \hline
$4$ & $0.125$ & $0.1875$ & $-0.0526$ & $-0.0079$ & $0.1270$ \\ \hline
$5$ & $0.143$ & $0.25$ & $-0.0789$ & $-0.0175$ & $0.1536$ \\ \hline
$6$ & $0.156$ & $0.3125$ & $-0.1052$ & $-0.0271$ & $0.1802$ \\ \hline
$7$ & $0.1\bar{6}$ & $0.375$ & $-0.1315$ & $-0.0367$ & $0.2068$ \\ \hline
$8$ & $0.175$ & $0.4375$ & $-0.1578$ & $-0.0463$ & $0.2334$ \\ \hline
\end{tabular}
\end{center}

\noindent {\small Table 3: n-particle contributions to the }$\Delta ${\small %
-theorem versus conformal dimensions in the $SU(N)_{2}/U(1)^{N-1}$ -WZNW
coset model.}

As we already observed for the c-theorem, the series converges slower for
larger values of $N$. The reason for this behaviour is the same, namely the
increasing symmetry factor. Note also that the next contribution is negative.

Following now the RG-flow for the conformal dimension (\ref{dr}) by varying $%
r_{0}$, we assume that the $\Delta (\lambda _{1},\lambda _{1})$-field flows
to the $\Delta (\lambda _{1},\lambda _{1})$-field in the corresponding new
cosets. Similar as for the Virasoro central charge we may compare the exact
expression 
\begin{eqnarray}
&&\Delta (\lambda _{1},\lambda _{1})_{SU(i+1)_{2}/U(1)^{i}\otimes
SU(N-i)_{2}/U(1)^{N-i-1}}= \\
&&\Delta (\lambda _{1},\lambda _{1})_{SU(N)_{2}/U(1)^{N-1}}+\frac{%
i(N+5)(N-i-1)}{4(N+2)(i+3)(N-i+2)}.  \nonumber
\end{eqnarray}
with the numerical results. The contributions (\ref{l1})-(\ref{l3}) yield 
\begin{eqnarray}
&&\lim_{\sigma _{i,i+1}\rightarrow \infty }{\small \Delta }^{\mu }{\small %
(\sigma }_{i,i+1}{\small ,\ldots )}= \\
&&{\small \Delta }^{\mu }{\small (\sigma }_{i,i+1}={\small 0,\ldots )+0.0359I%
}_{i,i+1}+{\small 0.0113I}_{i,i-1}.  \nonumber
\end{eqnarray}
Once again we find good agreement between the two computations for small
values of $N$. Our complete numerical results are presented in figure 3,
which confirm the outlined flow for various values of N.

\begin{center}
\includegraphics[width=8.2cm,height=6.09cm]{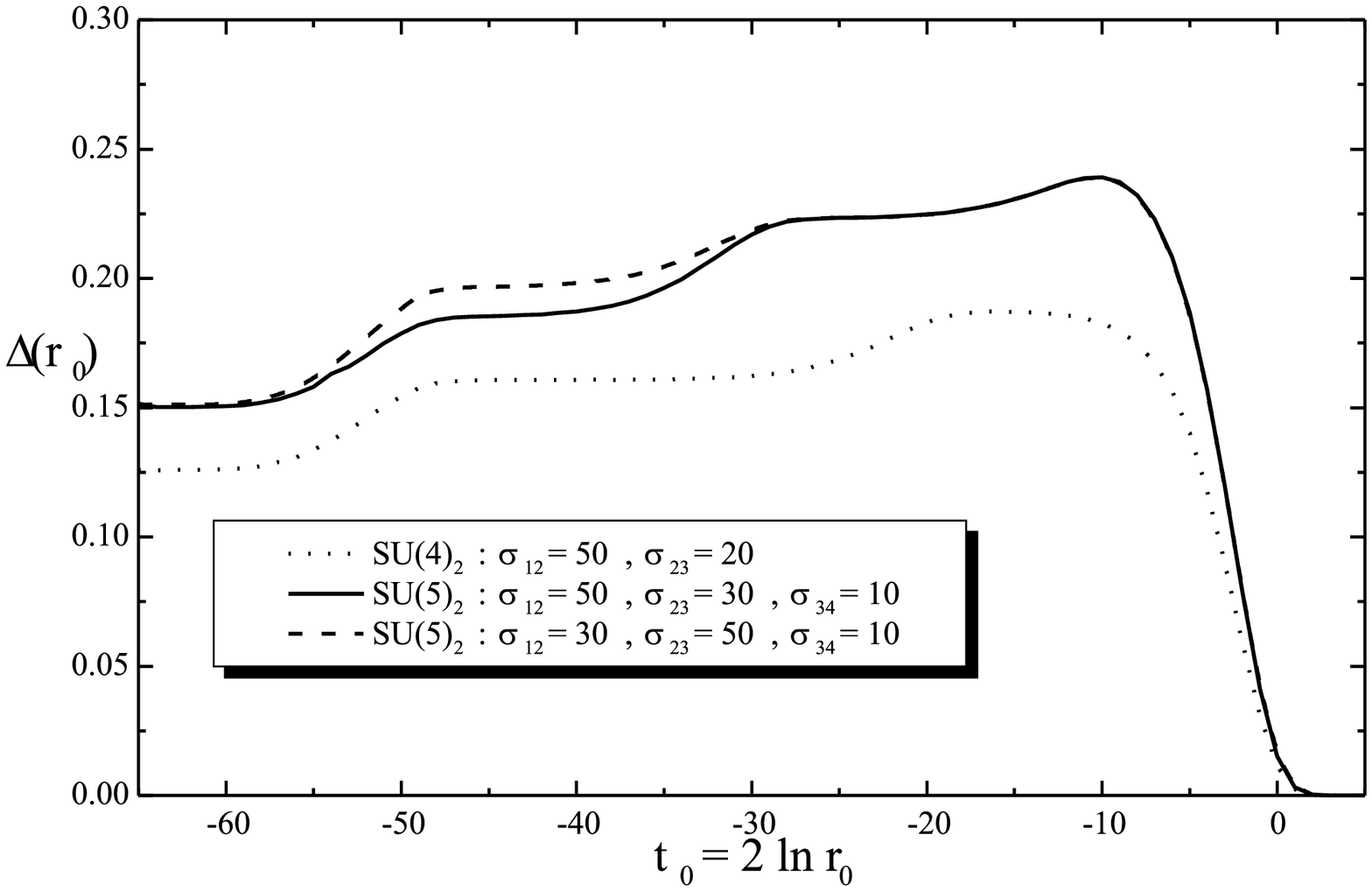}
\end{center}

\noindent {\small Figure 3: RG flow for the conformal dimension of $\mu$.}

\vspace*{1.2mm}

Notice by comparing the figures 3 and 1, that, as we expect, the transition
from one value for $\Delta $ to the one in the decoupled system occurs at
the same energy scale $t_{0}$ at which the value of the Virasoro central
charge flows to the new one.

In analogy to (\ref{CS}) we may now define a function ``$\beta ^{\prime }$''
and demand that it obeys the Callan-Symanzik equation 
\begin{equation}
r\frac{d}{dr}g^{\prime }=\beta ^{\prime }(g^{\prime })\,\,.  \label{betade}
\end{equation}

$\,\,$The ``coupling constant'' related to $\beta ^{\prime }$ is normalized
in such a way that it vanishes at the ultraviolet fixed point, i.e. $%
g^{\prime }:=\Delta (r)-\Delta _{\text{uv}}$, such that whenever we find $%
\beta ^{\prime }(\tilde{g}^{\prime })=0$, we can identify $\hat{\Delta}=%
\tilde{g}^{\prime }-\Delta _{\text{uv}}$ as the conformal dimension of the
operator under consideration of the corresponding conformal field theory.
From our analysis of (\ref{dr}) we may determine $\beta ^{\prime }$ as a
function of $g^{\prime }$ by means of (\ref{betade}). Our results are
presented in figure 4.

\begin{center}
\includegraphics[width=8.2cm,height=6.09cm]{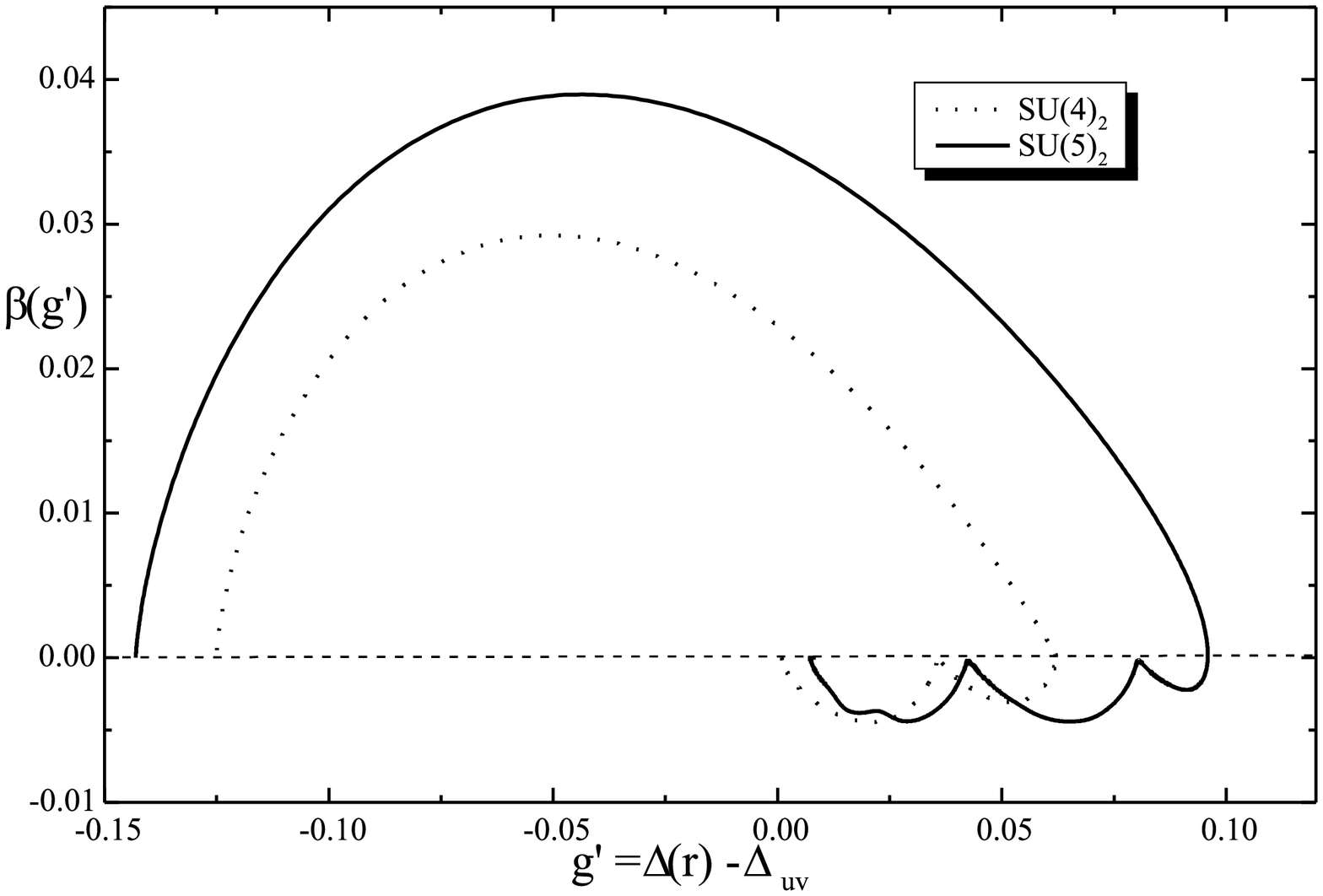}
\end{center}

\noindent {\small Figure 4: The $\beta ^{\prime}$-function .}

Once again, for $SU(4)_{2}$ the accuracy is very high and we clearly read
off from figure 4 the expected fixed points {\small \ }$\tilde{g}^{\prime
}=-0.125,0,0.0375,0.0625$. The $SU(5)_{2}$-fixed points $\tilde{g}^{\prime
}=-0.1429,0,0.0446,0.0821,0.1071$, are once again slightly shifted.

\begin{center}
\includegraphics[width=8.2cm,height=6.09cm]{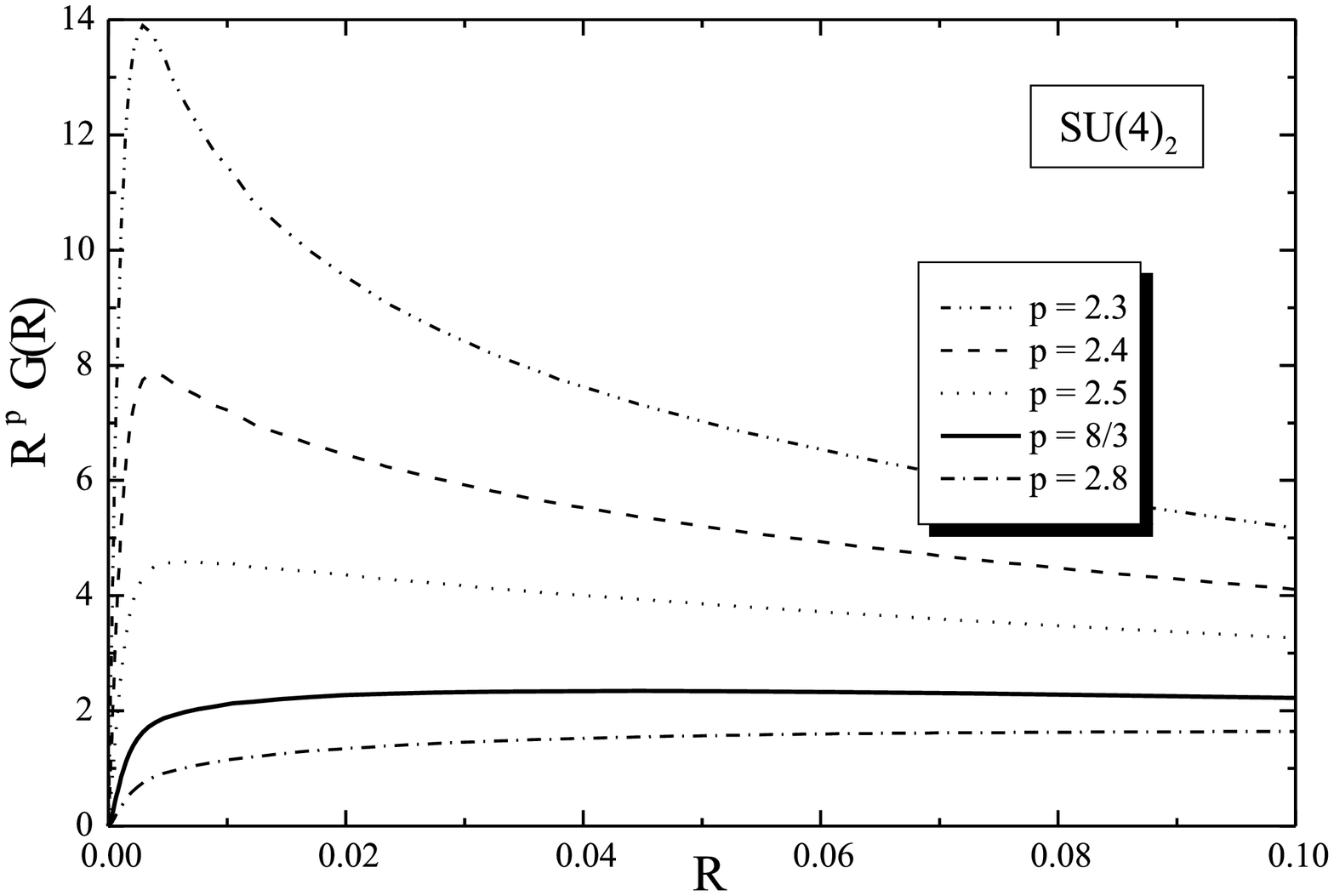}

\includegraphics[width=8.2cm,height=6.09cm]{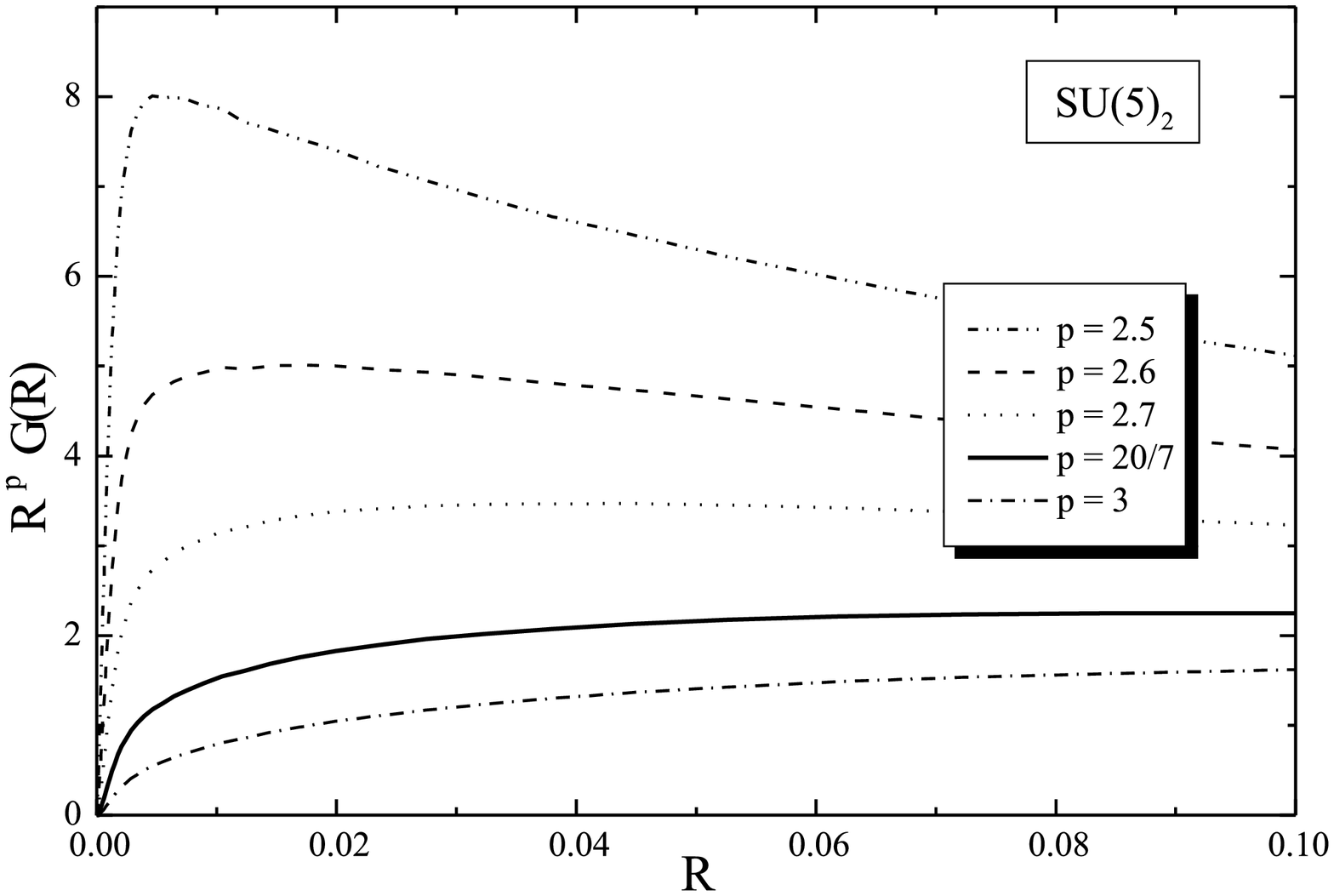}
\end{center}

\noindent {\small Figure 5: Rescaled correlation function G(R)=$\left\langle
\Theta (R)\Theta (0)\right\rangle \,$ as a function of $R=r m$.}

Unfortunately, whenever the correlation function between ${\cal O}$ and $%
\Theta $ is vanishing, or when we consider an operator which does not flow
to a primary field, we can not employ the delta sum rule (\ref{dr}).
Alternatively, we may exploit the well known relation

\begin{equation}
\lim_{r\rightarrow 0}\left\langle {\cal O}(r){\cal O}(0)\right\rangle \sim
r^{-4\Delta ^{\!\!{\cal O}}}.  \label{ultra}
\end{equation}
near the critical point in order to determine the conformal dimension. To
achieve consistency with the proposed physical picture we want to identify
in particular the conformal dimension of the perturbing operator. Recalling
that the trace of the energy momentum tensor is proportional to the
perturbing field we analyse $\left\langle \Theta (R)\Theta (0)\right\rangle $
for this purpose.

According to (\ref{ultra}), we deduce from figure 5 $\Delta =2/3,5/7$ for $%
N=4,5$, respectively, which coincides with the expected values.

\section{Conclusions}

One of the main deductions from our analysis is that the scattering matrix
proposed in \cite{HSGS} may certainly be associated to the perturbed gauged
WZNW-coset. This is based on the fact that we reproduce all the predicted
features of this picture, namely the expected ultraviolet Virasoro central
charge, various conformal dimensions of local operators and the
characteristics of the unstable particle spectrum.

Our construction of general solutions to the form factor consistency
equations certainly constitutes a further important step towards a generic
group theoretical understanding of the $n$-particle form factor expressions.
The next natural step is to extend the investigation towards higher level
algebras \cite{CF4}.

Concerning the computation of correlation functions, our results also
indicate that the ``folkloristic belief'' of the fast convergence of the
series expansion of (\ref{corre}) has to be challenged. In fact, for large
values of $N$, this is not true anymore. It would be highly desirable to
have more concrete quantitative criteria at hand.

Despite the fact of having identified some part of the operator content, it
remains a challenge to perform a definite one-to-one identification between
the solutions to the form factor consistency equations and the local
operators. It is clear that we require new additional technical tools to do
this, since the $\Delta $-sum rule (\ref{dr}) may not be applied in all
situations and (\ref{ultra}) does not allow a clear cut deduction of $\Delta 
$.

In comparison with other methods to achieve the same goal, we should note
that in principle we could obtain, apart from conformal dimensions different
from the one of the perturbing operator, the same qualitative picture from a
TBA-analysis \cite{CFKM}. However, in the latter approach the number of
coupled non-linear integral equations to be solved increases with $N$, which
means the system becomes extremely complex and cumbersome to solve even
numerically. Computing the scaling function with the help of form factors
only adds more terms to each $n$-particle contribution, but is technically
not more involved. The price we pay in this setting is, however, the slow
convergence of (\ref{corre}).

We conjecture that the ``cutting rule'' which describes the renormalization
group flow also holds for other groups. This is supported by the general
structure of the HSG-scattering matrix.\medskip

\noindent {\bf Acknowledgments: } A.F. is grateful to the Deutsche
Forschungsgemeinschaft (Sfb288) for financial support. O.A.C. thanks CICYT
(AEN99-0589), DGICYT (PB96-0960) and the Freie Universit\"{a}t Berlin for
partial financial support and is also very grateful to the Institut f\"{u}r
theoretische Physik of the Freie Universit\"{a}t for hospitality. We would
like to thank J.L. Miramontes for useful comments.

\begin{description}
\item  {\small \setlength{\baselineskip}{12pt}}
\end{description}

\end{document}